\documentclass[prl,aps,epsf,twocolumn,showpacs,10pt]{revtex4-1}
\usepackage{graphicx,amsfonts,times,bm,amsmath,verbatim,color,array}

\usepackage{subfigure}
\usepackage{float}
\usepackage{epsfig}
\usepackage{hyperref}
\hypersetup{
colorlinks=true,final=true,
        linkcolor=blue,
        citecolor=blue,
        filecolor=blue,
        urlcolor=blue,
}

\begin{document}
\title{Chiral anomaly as origin of planar Hall effect in Weyl semimetals}

\author{S. Nandy$^{1}$}
\author{Girish Sharma$^{2}$}
\author{A. Taraphder$^{1,3}$}
\author{Sumanta Tewari$^{1,4}$}

\affiliation{
$^{1}$ Department of Physics, Indian Institute of Technology Kharagpur, W.B. 721302, India\\
$^2$ Department of Physics, Virginia Tech, Blacksburg, VA 24061, U.S.A\\
$^{3}$ Centre for Theoretical Studies, Indian Institute of Technology Kharagpur, W.B. 721302, India\\
$^4$Department of Physics and Astronomy, Clemson University, Clemson, SC 29634, U.S.A}

\begin{abstract}
In condensed matter physics, the term ``chiral anomaly'' implies the violation of the separate number conservation laws of Weyl fermions of different chiralities in the presence of parallel electric and magnetic fields. One effect of chiral anomaly in the recently discovered Dirac and Weyl semimetals is a positive longitudinal magnetoconductance (LMC).
Here we show that chiral anomaly and non-trivial Berry curvature effects engender another striking effect in WSMs, the \textit{planar Hall effect} (PHE).
Remarkably, PHE manifests itself when the applied current, magnetic field, and the induced transverse ``Hall" voltage all lie in the same plane, precisely in a configuration in which the conventional Hall effect vanishes. In this work we treat PHE quasi-classically, and predict specific experimental signatures for type-I and type-II Weyl semimetals that can be directly checked in experiments.
\end{abstract}

\maketitle

\textit{Introduction:} In condensed matter physics the Weyl equation, originally introduced in high energy physics~\cite{Peskin}, describes the low energy quasiparticles near the touching of a pair of non-degenerate bands in a class of topological systems known as Weyl semimetals (WSM)
 ~\cite{Murakami1:2007, Murakami2:2007, Yang:2011, Burkov1:2011, Burkov:2011, Volovik, Wan:2011, Xu:2011}.
 In WSMs the momentum space touching points of non-degenerate pairs of bands act as source and sink of Abelian Berry curvature, an analog of magnetic field but defined in momentum space~\cite{Xiao:2010}. 
 WSMs violate spatial inversion and/or time reversal symmetries ~\cite{Wan:2011, Xu:2011, Burkov:2011, Volovik}, and are topologically protected by a non-zero flux of Berry curvature across the Fermi surface. 
 By Gauss's theorem, the flux of the Berry curvature known as Chern number is related to the strength of the magnetic monopole enclosed by the Fermi surface, and is quantized to integer values. It can be shown that ~\cite{Nielsen:1981, Nielsen:1983} in WSMs the Weyl points come in pairs of positive and negative monopole charges (also called chirality) and the net monopole charge summed over all the Weyl points in the Brillouin zone vanishes. 

In $\mathbf{k}\cdot\mathbf{p}$ theory, the effective Hamiltonian for the low energy linearly dispersing quasiparticles near an isolated Weyl point situated at momentum space point $\mathbf{K}$ can be written as,
\begin{eqnarray}
H_{\mathbf{k}} = \sum_{i=1}^{3} \; v_i (\mathbf{k}_i)\sigma_i,
\label{Eq_H_k_weyl_1}
\end{eqnarray}
where the crystal momenta $\mathbf{k}_i$ are measured from the band degeneracy point $\mathbf{K}$, $\hbar=c=1$, and $\sigma_i$s are the three Pauli matrices. 
WSMs evince many anomalous transport and optical properties, such as anomalous Hall effect in time reversal broken WSMs, dynamic chiral magnetic effect related to optical gyrotropy and natural optical activity in inversion broken WSMs \cite{Goswami:2015,Zhong}, and, most importantly, negative longitudinal magnetoresistance in the presence of parallel electric and magnetic fields due to non-conservation of separate electron numbers of opposite chirality for relativistic massless fermions, an effect known as the chiral or Adler-Bell-Jackiw anomaly~\cite{Goswami:2015,Zhong,Goswami:2013,Adler:1969, Bell:1969, Nielsen:1981, Nielsen:1983, Aji:2012, Zyuzin:2012, Volovik, Wan:2011, Xu:2011, Goswami:2015, Goswami:2013}.

In the absence of parallel electric and magnetic fields in WSMs, as for relativistic chiral fermions in high energy physics, the numbers of right and left handed Weyl fermions (i.e. Weyl fermions of different chiralities) are separately conserved. However, in the presence of externally imposed parallel electric and magnetic fields,
the separate number conservation laws are violated~\cite{Adler:1969, Bell:1969}, leaving only the total number of fermions to be conserved.
For weak magnetic fields for which the Landau level quantization is wiped out by disorder effects, a semiclassical description~\cite{Son:2013, Kim:2014} of magnetoresistance suggests that $\mathbf{E}\cdot\mathbf{B} \neq 0$ leads to a positive LMC as a result of chiral anomaly, while the transverse magnetoresistance remains positive and conventional. Consistent with this picture, recently, several experimental groups have found the evidence of chiral anomaly induced positive LMC in Dirac and Weyl materials~\cite{He:2014, Liang:2015, CLZhang:2016, QLi:2016, Xiong,Hirsch}.

In this paper we discuss a second effect of chiral anomaly, the so-called \textit{planar} Hall effect~\cite{Burkov:2017}, i.e. appearance of an \textit{in-plane} transverse voltage when the \textit{co-planar} electric and magnetic fields are not perfectly aligned to each other. The planar Hall conductivity $\sigma_{yx}$, i.e. the transverse conductivity measured across the $\hat{y}$ direction perpendicular to the applied electric field and current in the $\hat{x}$ direction in the presence of a magnetic field in $x-y$ plane making an angle $\theta$ with the $x$ axis, is known to occur in ferromagnetic systems~\cite{KY:1968, Dobrowolska:2007, Bowen_2005, Keizer_2007, Friedland_2006} with dependence on $\theta$ similar to what we find here for WSMs. It has also been observed recently with similar angular dependence in the surface state of a topological insulator where it has been linked to magnetic field induced anisotropic lifting of the protection of the surface state from backscattering~\cite{Taskin:2017}. Here we develop a quasi-classical theory of planar Hall effect in Weyl semimetals, where the electron or hole Fermi surfaces enclose non-zero fluxes of Berry curvature in momentum space.
Unlike anomalous Hall effect understood quasi-classically in terms of Berry curvature effects \cite{Jung}, to the best of our knowledge planar Hall effect has so far not been discussed as a topological response function. Our treatment of PHE in terms of chiral anomaly and non-trivial Berry curvature, along with specific experimental signatures in type-I and type-II WSMs, is an important first step to fill this gap.

\textit{Model Hamiltonian:}
The momentum space Hamiltonian for a generic single chiral Weyl node can be expressed as
\begin{equation}
{\cal H_{\mathbf{k}}}^{\chi}={\hbar}v_{F}(\chi\mathbf{k}\cdot\boldsymbol{\sigma}+Ck_{x}\sigma_{0})
\label{H_linear}
\end{equation}
where $v_{F}$ is the Fermi velocity, $\chi$ is the chirality associated with the Weyl node, $\boldsymbol{\sigma}$ represent the vector of Pauli matrices, $\sigma_{0}$ is the identity matrix, and $C$ is the tilt parameter which can be taken along the $k_{x}$ direction without any loss of generality~\cite{Soluyanov:2015}.
When the anisotropy is zero i.e. $C=0$, electron and hole bands touch at the Weyl point leading to a point like Fermi surface. When the anisotropy along $k_{x}$ is small enough ($C=0.5$), the Fermi surface is still point-like and is classified as the type-I Weyl node. With the increase in anisotropy ($|C|$ $>$ 1), electron and hole pockets now appear at the Fermi surface leading to a distinct phase which is classified as a type-II Weyl node. 
\begin{figure}[htb]
\begin{center}
\epsfig{file=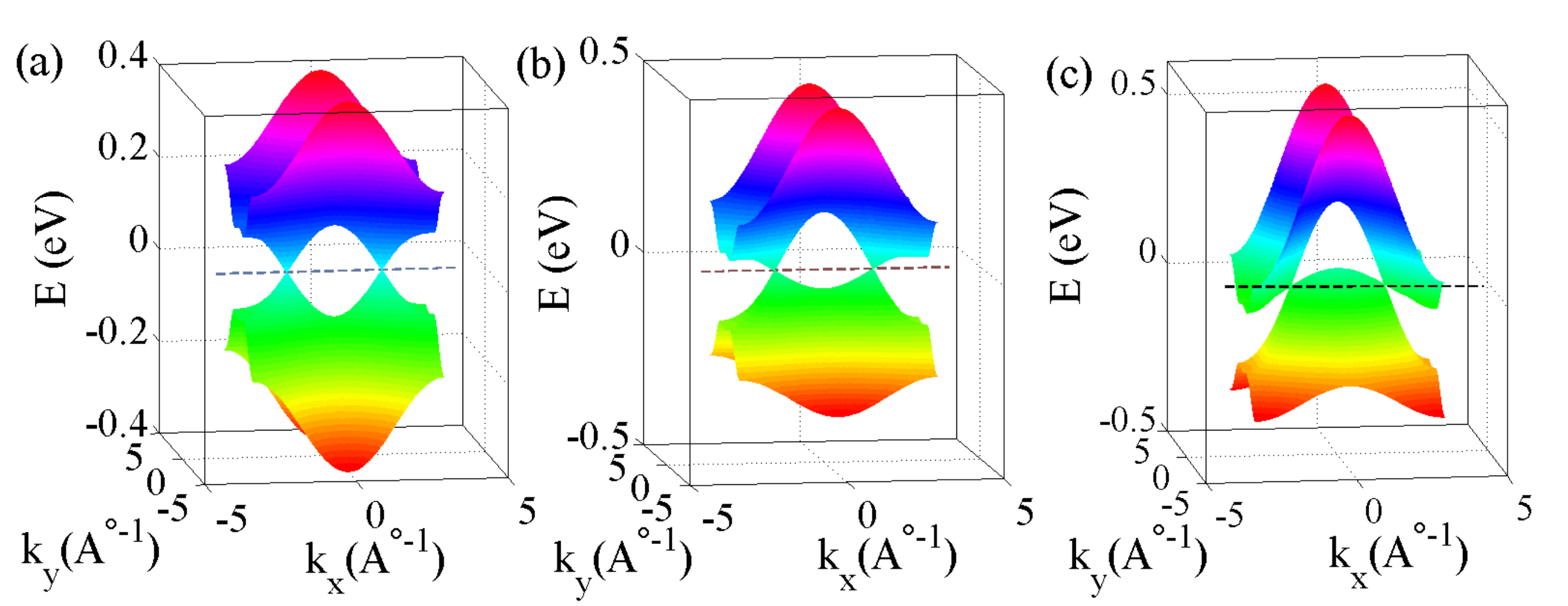,trim=0.0in 0.05in 0.0in 0.05in,clip=true, width=90mm}\vspace{0em}
\caption{(Color online) 3D band dispersion of the lattice model of Weyl semimetal ($k_{z}$ is suppressed) obtained by diagonalizing Hamiltonian in Eq.~(\ref{H_total}) for (a) $\gamma=0$ (type-I Weyl nodes), (b) $\gamma=0.05$ (type-I Weyl nodes) and (c) $\gamma=0.15$ (type-II Weyl nodes) respectively. The chemical potential is set at zero energy (indicated by dash line). The Weyl cones are at ($k_{0}$,0,0) and (-$k_{0}$,0,0). The parameters used are $t=-0.005$ eV, $m=0.15$ eV, and $k_{0}=\frac{\pi}{2}$.}
\label{node_dis}
\end{center}
\end{figure}

\textit{Planar Hall effect:}
We will now investigate the electronic contributions to the planar Hall conductivity in the quasi-classical Boltzmann formalism \cite{Son:2013,Kim:2014}. The Boltzmann formalism is valid because for the scattering time $\tau \sim 10^{-13}$ s in typical Dirac and Weyl semimetals~\cite{XiongEPL,Watzman_2017, ShekharNature}, and the effective mass $m^*\sim 0.11 m_e$~\cite{XiongEPL,ShekharNature} , $\omega_c\tau \sim 0.3 < 1$ for typical magnetic field $B\sim 3-5$ T, where $\omega_c = eB/m^*c$ is the cyclotron frequency. Additionally, we use the standard relaxation time approximation \cite{Son:2013,Kim:2014} which assumes that any perturbation in the system decays exponentially with a characteristic time constant $\tau$. This approximation is valid in isotropic systems with elastic (impurity dominated) scattering processes generally valid in WSMs \cite{Lundgren:2014}.
We begin with the linear response equation for the charge current ($\mathbf{J}$) to external perturbative fields (electric field $\mathbf{E}$ and temperature gradient $\nabla T$), which is given by
\begin{align}
J_{a}=\sigma_{ab}E_{b}+\alpha_{ab}(-\nabla_{b} T)
\label{e1}
\end{align}
where $\hat{\sigma}$ and $\hat{\alpha}$ are different conductivity tensors.
A phenomenological Boltzmann transport equation can be written as~\cite{Ziman}
\begin{equation}
\left(\frac{\partial}{\partial t}+\mathbf{\dot{r}}\cdot\mathbf{\nabla_{r}}+\mathbf{\dot{k}}\cdot\mathbf{\nabla_{k}}\right)f_{\mathbf{k},\mathbf{r},t}=I_{coll} \{f_{\mathbf{k},\mathbf{r},t}\}
\label{e3}
\end{equation}
where on the right side $I_{coll} \{f_{\mathbf{k},\mathbf{r},t}\}$ is the collision integral which incorporates the effects of electron correlations and impurity scattering. We are interested in computing the electron distribution function which is given by $f_{\mathbf{k},\mathbf{r},t}$.
Since we are primarily interested in steady-state solutions to the Boltzmann equation, Eq.~(\ref{e3}) can be rewritten as
\begin{equation}
(\mathbf{\dot{r}}\cdot\mathbf{\nabla_{r}}+\mathbf{\dot{k}}\cdot\mathbf{\nabla_{k}})f_{k}=\frac{f_{eq}-f_{\mathbf{k}}}{\tau(\mathbf{k})}
\label{e5}
\end{equation}
where we have invoked the relaxation time approximation (RTA) for the collision integral and also dropped the $\mathbf{r}$ dependence of $f_{\mathbf{k},\mathbf{r},t}$, valid for spatially uniform fields. 
The relaxation time $\tau (\mathbf{k})$ on the Fermi surface can in general have a momentum dependence but we will ignore this dependence in our work as it doesn't change any of our qualitative conclusions. The function $f_{eq}$ is the equilibrium Fermi-Dirac distribution function which describes electron distribution in the absence of any external fields.  
It is now well established that the low energy transport properties are substantially modified due to the Berry curvature of the electron wave functions~\cite{Xiao:2010}.

To calculate planar Hall effect, we apply an electric field ($\mathbf{E}$) along the $x-$axis and a magnetic field ($\mathbf{B}$) in the $xy$ plane at a finite angle $\theta$ from the $x-$axis, i.e. $\mathbf{B}=B\cos\theta\hat{x}+B\sin\theta\hat{y}$, $\mathbf{E}=E\hat{x}$.
 In the presence of Berry curvature associated with a single chiral Weyl node, the quasi-classical equations of motion are~\cite{Duval:2006}
\begin{align}
\mathbf{\dot{r}}=D(\mathbf{B,\Omega_{k}})[\mathbf{v_{k}}+\frac{e}{\hbar}(\mathbf{E}\times\mathbf{\Omega_{k}})+\frac{e}{\hbar}(\mathbf{v_{k}}\cdot\mathbf{\Omega_{k}})\mathbf{B}]\label{e6}\\
\hbar\mathbf{\dot{k}}=D(\mathbf{B,\Omega_{k}})[e\mathbf{E}+\frac{e}{\hbar}(\mathbf{v_{k}} \times \mathbf{B})+\frac{e^{2}}{\hbar}(\mathbf{E}\cdot\mathbf{B})\mathbf{\Omega_{k}}]
\label{e7}
\end{align}
Here, $D(\mathbf{B,\Omega_{k}})=(1+\frac{e}{\hbar}(\mathbf{B}.\mathbf{\Omega_{k}}))^{-1}$ is the phase space factor, where $\mathbf{\Omega_{k}}$ is the Berry curvature, and $\mathbf{v_{k}}$ is the group velocity ~\cite{Niu:2006}. For ease of notation hereafter we will simply denote $D(\mathbf{B,\Omega_{k}})$ by $D$, dropping the implied $\mathbf{B}$ and $\mathbf{\Omega_k}$ dependence.
Substituting the above equations of motion into the steady state Boltzmann equation Eq.~(\ref{e5}), it then takes the form
\begin{align}
&(\frac{eEv_{x}}{\hbar}+\frac{e^{2}}{\hbar}BE\cos\theta(\mathbf{v_{k}}.\mathbf{\Omega_{k}}))\frac{\partial f_{eq}}{\partial \epsilon}+\frac{eB}{\hbar^{2}}(-v_{z}\sin\theta\frac{\partial}{\partial k_{x}} \nonumber \\
&(v_{x}\sin\theta-v_{y}\cos\theta)\frac{\partial}{\partial k_{z}}+v_{z}\cos\theta\frac{\partial}{\partial k_{y}})f_{\mathbf{k}} =\frac{f_{eq}-f_{\mathbf{k}}}{D\tau}
\label{e9}
\end{align}
We solve the above equation by assuming the following ansatz for the deviation of the electron distribution function $\delta f_{\mathbf{k}}=f_{\mathbf{k}}-f_{eq}$
\begin{align}
\delta f_{\mathbf{k}}=(eDE\tau{v_{x}}+&\frac{e^{2}DBE\tau\cos \theta (\mathbf{v_{k}}\cdot\mathbf{\Omega_{k}})}{\hbar}+\mathbf{v}\cdot\mathbf{\Gamma})\left(\frac{\partial f_{eq}}{\partial\epsilon}\right)
\label{e10}
\end{align}
where $\mathbf{\Gamma}$ is correction factor due to magnetic field $\mathbf{B}$. Plugging $\delta f_{\mathbf{k}}$ into Eq.~(\ref{e9}), we have
\begin{align}
&\frac{eB}{\hbar^{2}}\left(-v_{z}\sin\theta\frac{\partial}{\partial k_{x}}+v_{z}\cos\theta\frac{\partial}{\partial k_{y}}+(v_{x}\sin\theta-v_{y}\cos\theta)\frac{\partial}{\partial k_{z}}\right) \nonumber \\
&(eED\tau(v_{x}+\frac{eB\cos\theta}{\hbar}(\mathbf{v_{k}}\cdot\mathbf{\Omega_{k}}))+\mathbf{v}\cdot\mathbf{\Gamma})=\frac{\mathbf{v}\cdot\mathbf{\Gamma}}{D\tau}
\label{e11}
\end{align}
We now calculate the correction factor $\mathbf{\Gamma}$ which vanishes in the absence of any magnetic field $\mathbf{B}$ by expanding the inverse band-mass which arises in Eq.~(\ref{e11}), and noting the fact that the above equation is valid for all values of velocity.
The Boltzmann distribution function $f_{\mathbf{k}}$ is then evaluated to be,
\begin{align}
f_{\mathbf{k}}&=f_{eq}-eDE\tau({v_{x}}+\frac{eB\cos \theta}{\hbar}(\mathbf{v_{k}}\cdot\mathbf{\Omega_{k}}))\frac{\partial f_{eq}}{\partial\epsilon} \nonumber \\
&-eDE\tau(v_{x}c_{x}\sin \theta+v_{y}c_{y}\cos \theta+v_{z}c_{z}))\frac{\partial f_{eq}}{\partial\epsilon} \nonumber \\
\label{e16}
\end{align}
where $c_x$, $c_y$ and $c_z$ are correction factors which incorporate Berry phase effects and are related to $\mathbf{\Gamma}$ (see supplementary material).

In the absence of any thermal gradient, the charge current can be written as
$\mathbf{J}=e\int{[d^{3}k]}D^{-1}\mathbf{\dot{r}}f_{\mathbf{k}}$, accounting for the modified density of states due to the phase space factor $D$.
\begin{figure}[htp]
\centering
\begin{tabular}{cc}
\includegraphics[width=44mm]{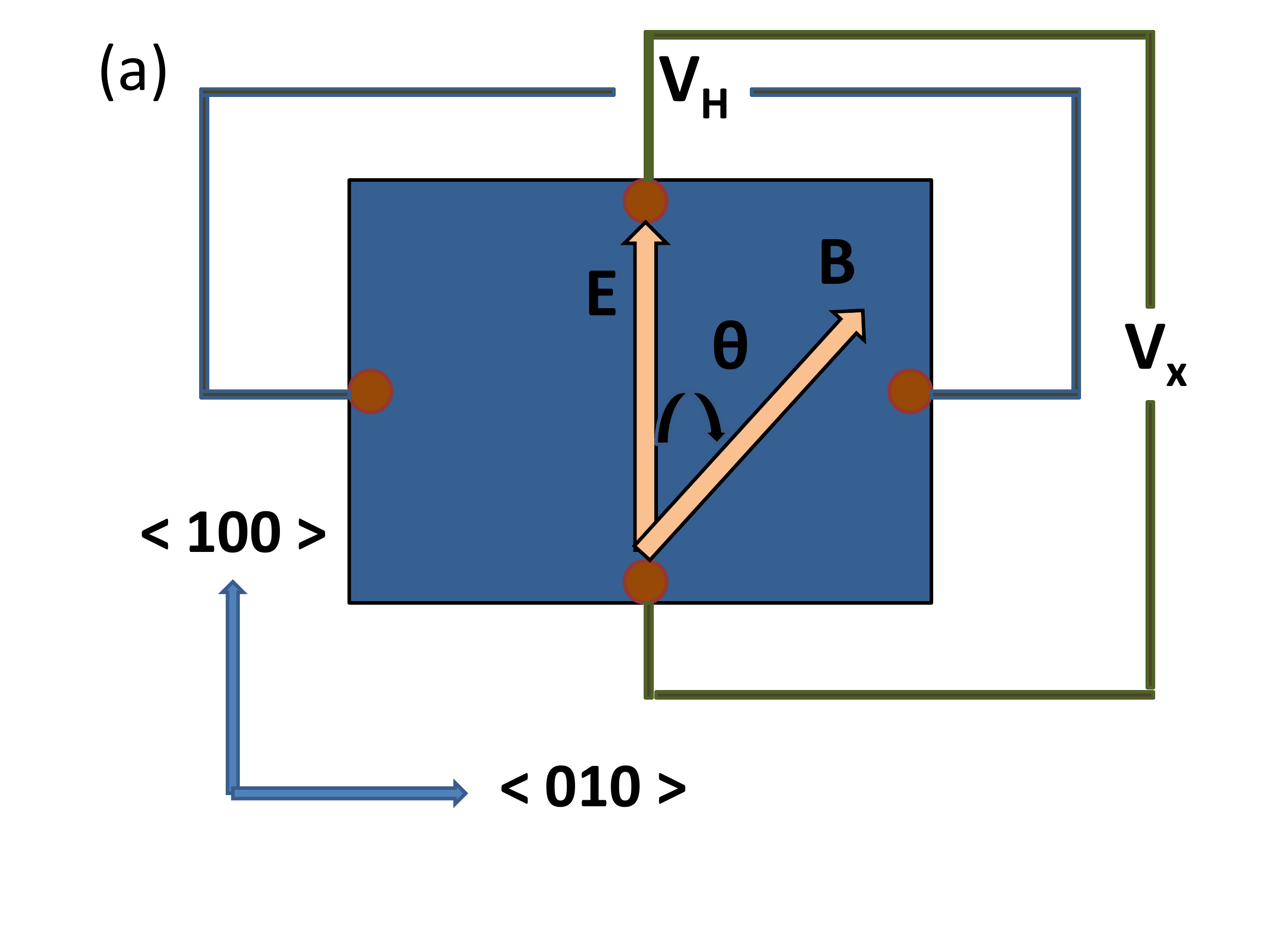}
\includegraphics[width=46mm]{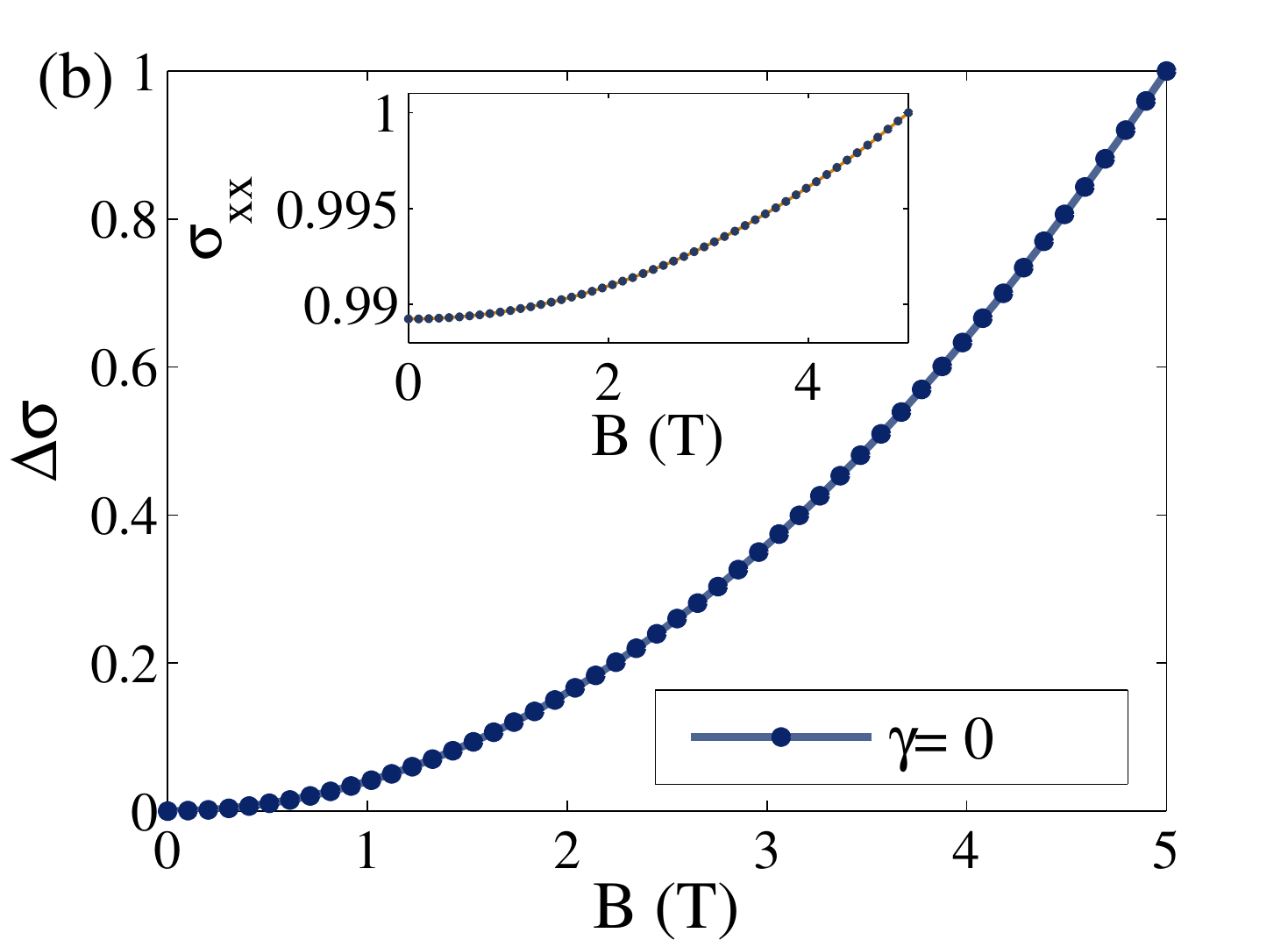}\\
\includegraphics[width=88mm]{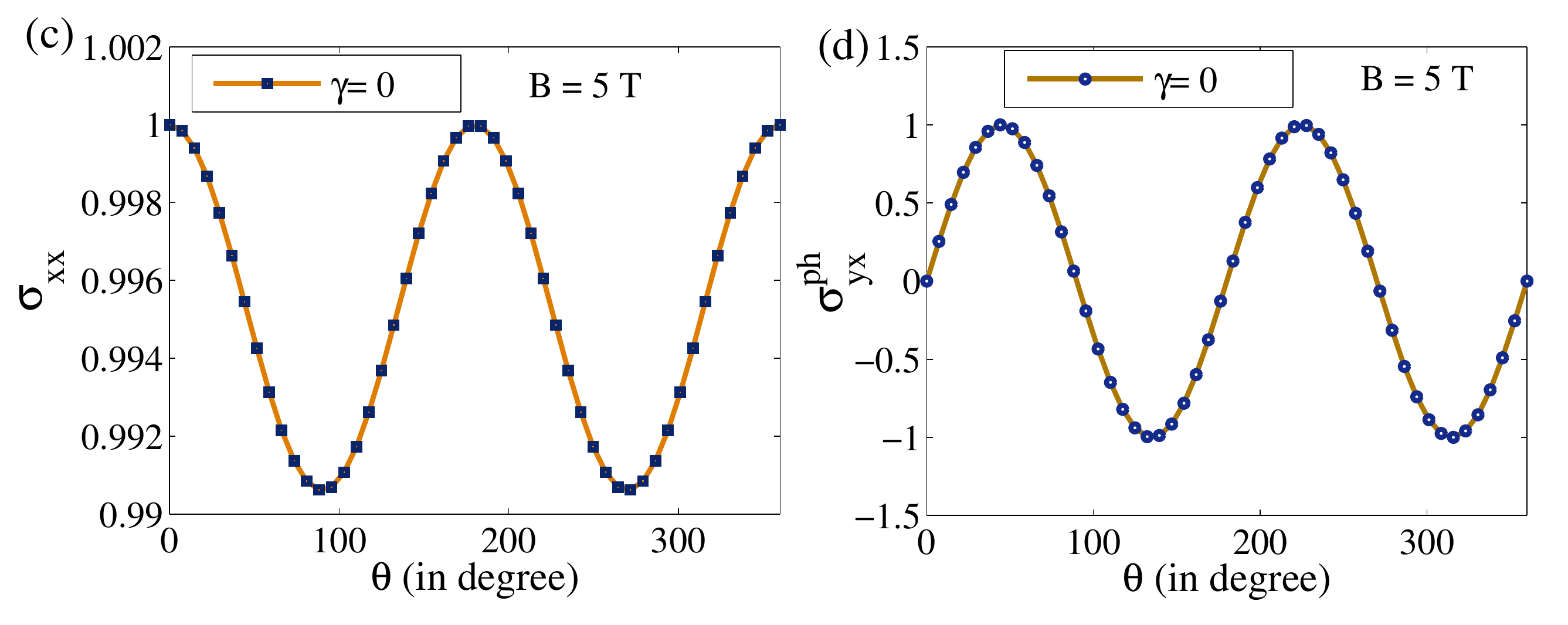}
\end{tabular}
\caption{(Color online) (a) Illustration for planar Hall effect measurement set-up ($V_{H}$ is the in-plane Hall voltage). (b) shows the normalized amplitude of planar Hall conductivity($\Delta\sigma$) as a function of magnetic field for the lattice model of Weyl semimetal given by the Eq.~(\ref{H_total}) for $\gamma=0$. (Inset shows Longitudinal magneto-conductivity as a function of magnetic field). Both PHC and LMC amplitudes show $B^2$ dependence on magnetic field. (c)-(d) depict the angular dependence of longitudinal magneto-conductivity and planar Hall conductivity for $B=5$ T. (We have normalized the $y$-axes of (c) and (d) by $\sigma_{xx}(\theta=0)$ and $\sigma_{yx}^{\text{ph}}(\theta=\pi/4)$ respectively.)}
\label{Weyl_1}
\end{figure}
Substituting $f_{k}$ into this equation and comparing it with Eq.~(\ref{e1}), we now arrive at the expression for the longitudinal electrical conductivity
\begin{align}
\sigma_{xx}&=e^{2}\int\frac{d^{3}k}{(2\pi)^{3}}\tau [D({v_{x}}+\frac{eB\cos \theta}{\hbar}(\mathbf{v_{k}}\cdot\mathbf{\Omega_{k}}))^{2}  ]\left(-\frac{\partial f_{eq}}{\partial \epsilon}\right)
\label{e19}
\end{align}
where we have dropped the other terms which vanish upon integration around a single Weyl node, or are of a much smaller order of magnitude compared to others in typical Weyl metals. In the above equation the anomalous velocity factor $\frac{eB\cos\theta}{\hbar}(\mathbf{v_{k}}\cdot\mathbf{\Omega_{k}})$ arises due to the topological chiral anomaly term which gives a finite $\mathbf{B}-$dependent longitudinal electrical conductivity, which is otherwise absent for a regular Fermi liquid. 
When $\theta=0$, we recover the formula for LMC for parallel $\mathbf{E}$ and $\mathbf{B}$ fields as derived in earlier works~\cite{Son:2013,Kim:2014, Lundgren:2014, Sharma:2016}. Now substituting $f_{k}$ from Eq.~(\ref{e16}) into Eq.~(\ref{e1}), we then arrive at the following expression for the electrical Hall conductivity
\begin{align}
\sigma_{yx}&=e^{2}\int\frac{d^{3}k}{(2\pi)^{3}}D\tau\left(-\frac{\partial f_{eq}}{\partial \epsilon}\right) [(v_{y}+\frac{eB\sin \theta}{\hbar}(\mathbf{v_{k}}\cdot\mathbf{\Omega_{k}})) \nonumber \\
&(v_{x}+\frac{eB\cos \theta}{\hbar}(\mathbf{v_{k}}\cdot\mathbf{\Omega_{k}}))]-\frac{e^{2}}{\hbar}\int\frac{d^{3}k}{(2\pi)^{3}}\Omega_{z}f_{eq}+e^{2} \nonumber \\ &\int\frac{d^{3}k}{(2\pi)^{3}}\tau
(\sin \theta c_{x}v_{x}+\cos \theta c_{y}v_{y}+c_{z}v_{z})v_{y}\left(-\frac{\partial f_{eq}}{\partial \epsilon}\right)
\label{e20}
\end{align}
\begin{figure}[htb]
\begin{center}
\epsfig{file=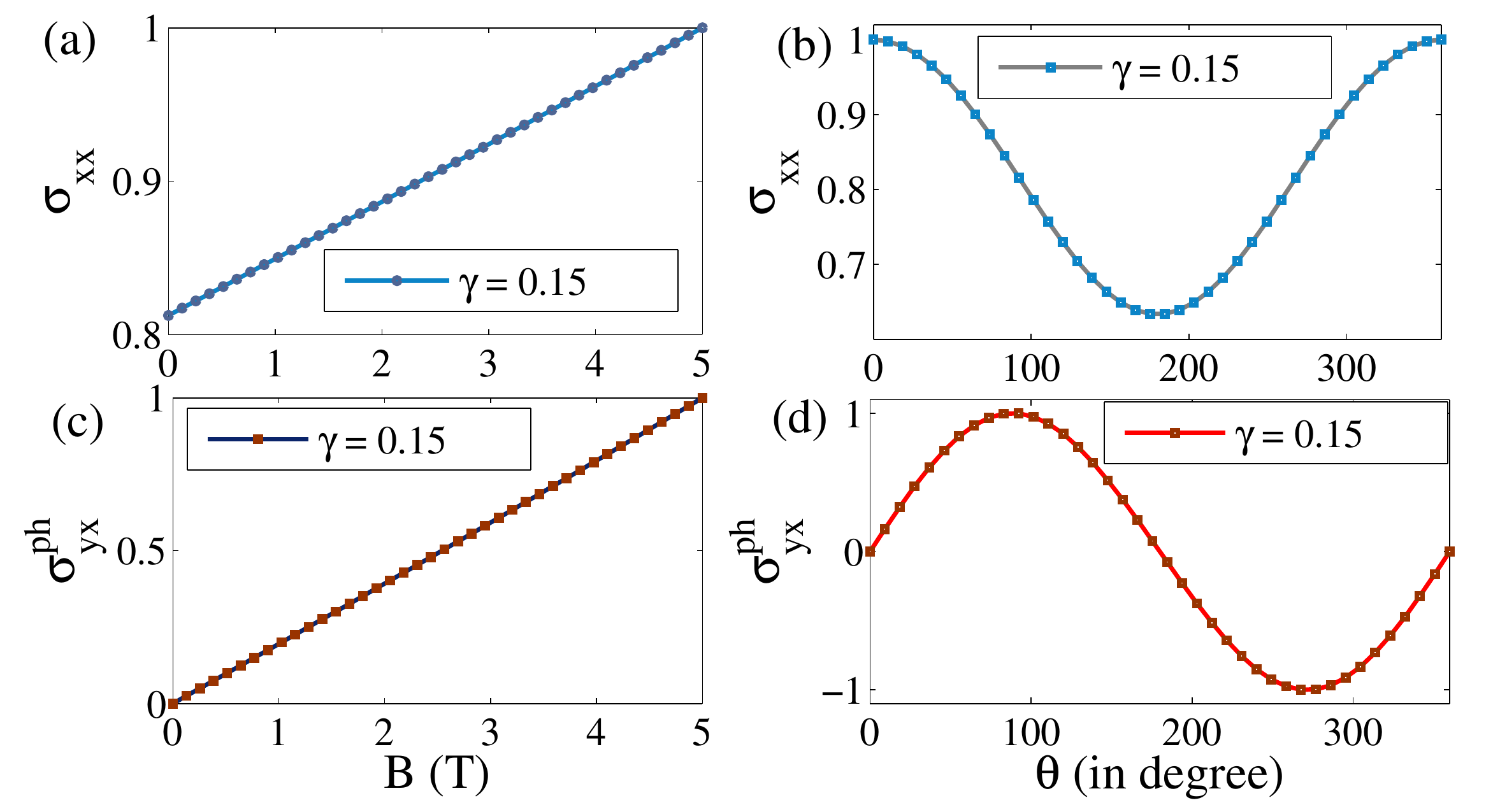,trim=0.0in 0.05in 0.0in 0.05in,clip=true, width=85mm}\vspace{0em}
\caption{(Color online) (a) shows the normalized longitudinal magneto-conductivity computed numerically for the lattice model of Weyl semimetal given by the Eq.~(\ref{H_total}) with tilt parameter $\gamma=0.15$ as a function of the magnetic field $\mathbf{B}$ applied along the tilt direction ($x$-axis). (b) depicts the angular dependence of longitudinal magneto-conductivity for $B=5$ T (applied parallel to tilt direction) and $\gamma=0.15$. (c)-(d) depict the same for planar Hall conductivity for the parameters mentioned above.}

\label{WSM_2}
\end{center}
\end{figure}
In the above expression the second momentum space integral (of the Berry curvature $\Omega_z$) in the above equation corresponds to the regular anomalous Hall contribution ($\sigma_{xy}^\text{a}$) from a single Weyl node.
Summed over all the nodes this term is non-zero for time reversal broken WSMs but vanishes for inversion broken WSMs as the integral over the Berry curvature vanishes in the presence of time reversal symmetry. We shall not consider this term any further as we are only interested in the chiral anomaly induced contribution to the Hall conductivity. We also note that our present Boltzmann treatment with energy independent scattering time defined on the Fermi surface is valid for $\mu  >>  k_{B}T, \hbar \omega_{c}$~\cite{Kim:2014,Lundgren:2014}, and in this limit the values of the terms involving $c_x$, $c_y$, $c_z$ are orders of magnitude smaller than the contribution from the rest in Eq.~(\ref{e20}).
We then arrive at our final expression for the chiral anomaly induced planar Hall conductivity,
\begin{align}
\sigma_{yx}^{\text{ph}}&=e^{2}\int\frac{d^{3}k}{(2\pi)^{3}}D\tau\left(-\frac{\partial f_{eq}}{\partial \epsilon}\right) [\frac{eB\sin \theta}{\hbar}(\mathbf{v_{k}}\cdot\mathbf{\Omega_{k}}) \nonumber \\
&(v_{x}+\frac{eB\cos \theta}{\hbar}(\mathbf{v_{k}}\cdot\mathbf{\Omega_{k}}))]
\label{e21}
\end{align}
where the superscript `ph' stands for ``planar Hall'' effect.

Eqs.~({\ref{e19},\ref{e21}}) are the central results of this paper. The numerical calculations to compute LMC and PHC have been performed for a prototype lattice model of a time reversal symmetry breaking Weyl semimetal with the lattice regularization providing a physical ultra-violet  cut-off to the momentum integrals. The prototype lattice model is given by,
\begin{equation}
{\cal H_{\mathbf{k}}}={H^{L}(\mathbf{k})}+{H^{T}(\mathbf{k})}
\label{H_total}
\end{equation}
where $H^{L}$ produces a pair of Weyl nodes of type-I at ($\pm k_{0}$,$0$,$0$)~\cite{Nandini_2017},
\begin{eqnarray}
{H^{L}(\mathbf{k})}&=(m(\cos(k_{y}b)+\cos(k_{z}c)-2)+2t(\cos(k_{x}a)\nonumber \\
&-\cos k_{0}))\sigma_{1}-2t\sin(k_{y}b)\sigma_{2}-2t\sin(k_{z}c)\sigma_{3} \nonumber \\
\label{H_lattice}
\end{eqnarray}
Here, m is the mass and t is hopping parameter.
The second term of the Hamiltonian $H^{T}$ tilts the nodes along $k_{x}$ direction, and can be written as,
\begin{equation}
{H^{T}(\mathbf{k})}=\gamma(\cos(k_{x}a)-\cos k_{0})\sigma_{0}
\label{H_tilt}
\end{equation}
where $\gamma$ is the tilt parameter. We first examine Eq.~({\ref{e21}}) for $\gamma=0$, the case of a  type-I WSM. After performing the momentum space integrals, and retaining only the non-vanishing terms, $\sigma^{\text{ph}}_{xy}$ is given by
\begin{align}
\sigma_{yx}^{\text{ph}}&=e^{2}\int\frac{d^{3}k}{(2\pi)^{3}}D\tau\left(-\frac{\partial f_{eq}}{\partial \epsilon}\right) \frac{e^2B^2\sin \theta \cos \theta}{\hbar^2}(\mathbf{v_{k}}\cdot\mathbf{\Omega_{k}})^{2}
\label{sxy_chiral}
\end{align}
Clearly, when $\theta=0,\pi/2$, $\sigma_{yx}^{\text{ph}}=0$, as expected, and the net Hall conductivity is determined by the Berry phase induced anomalous Hall contribution (if present as in the case of a time reversal broken WSM). But $\sigma_{yx}^{\text{ph}}$ in Eq.~(\ref{sxy_chiral}) is generically non-zero for any other arbitrary angle.

Using Eqs.~({\ref{e19},\ref{e21}}) we can now express $\sigma_{xx}$ and $\sigma_{yx}^{\text{ph}}$ in terms of the diagonal components of the conductivity tensor, $\sigma_{\parallel}$ and $\sigma_{\perp}$, corresponding to the cases when the current flows along and perpendicular to the magnetic field. Substituting $\theta=0$ and $\theta=\pi/2$ into Eq.~(\ref{e19}), we have
\begin{align}
&\sigma_{\parallel}=\sigma+e^{4}\int\frac{d^{3}k}{(2\pi)^{3}}D\tau\left(-\frac{\partial f_{eq}}{\partial \epsilon}\right) \frac{B^2}{\hbar^2}(\mathbf{v_{k}}\cdot\mathbf{\Omega_{k}})^{2} \nonumber \\
&\sigma_{\perp}=\sigma
\label{s_para}
\end{align}
Eq.~(\ref{e19}) and Eq.~(\ref{sxy_chiral}) thus take the form
\begin{align}
&\sigma_{xx}=\sigma_{\perp}+\Delta\sigma \cos^{2} \theta \nonumber \\
&\sigma_{yx}^{\text{ph}}=\Delta\sigma \sin \theta \cos \theta
\label{sxx_chiral}
\end{align}
where $\Delta\sigma=\sigma_{\parallel}-\sigma_{\perp},$ gives the anisotropy in conductivity due to chiral anomaly. The amplitude of planar Hall conductivity shows  $B^{2}$-dependence i.e. $\Delta\sigma \propto B^{2}$ for any value of $\theta$ except for $\theta=0$ and $\theta=\pi/2$ as shown in Fig.~\ref{Weyl_1}(b) whereas LMC has the finite value for all field directions and follows the $B^{2}$ dependence except at $\theta=\pi/2$ (Inset of Fig.~\ref{Weyl_1}(b)).
The longitudinal magnetoconductivity has the angular dependence of $\cos^{2} \theta$ which is shown in Fig.~\ref{Weyl_1}(c), leading to the anisotropic magnetoresistance(AMR)~\cite{Pan, Hong:1995, Tang:2003} whereas the planar Hall conductivity follows the $\cos \theta \sin \theta$ dependence as depicted in Fig.~\ref{Weyl_1}(d). Note that the planar Hall conductivity discussed here does not satisfy the antisymmetry property of regular Hall conductivity $(\sigma_{xy}\rho_{yx}=-1)$ since its origin is linked to the topological chiral anomaly term and not to a conventional Lorentz force and this fact can be used to remove the regular Hall contribution from the total Hall response to isolate PHE in experiments by taking measurements with both positive and negative B. On the other hand, the planar Hall effect can be distinguished from anomalous Hall effect by taking measurements with both $B=0$ and $B\ne 0$ and subtracting the background ($B=0$) contribution.


In Fig.~\ref{WSM_2} we have plotted the numerically calculated LMC ($\sigma_{xx}$) for a type-II WSM as a function of $\mathbf{B}$, where $\mathbf{E}$ is applied along the tilt direction ($x$ axis). Our calculations suggest that the LMC follows a $\mathbf{B}$-linear dependence~\cite{Sharma2:2016} when both applied magnetic field and electric field are parallel to the tilt axis (also valid for $0 \le \theta$ $<$ $\pi/2$ in first quadrant of the plane Fig.~\ref{Weyl_1}(a)) as depicted in Fig.~\ref{WSM_2}(a). For non-zero magnetic field, LMC shows $\cos \theta$ dependence for the same configuration of the applied $\mathbf{E}$ and $\mathbf{B}$ as shown in Fig.~\ref{WSM_2}(b). Further the planar Hall conductivity ($\sigma_{yx}^{\text{ph}}$ at $\theta=\pi/4$) computed using Eq.~(\ref{e21}) also follows the linear $\mathbf{B}$ dependence at any angle $0$ $<$ $\theta$ $\le$ $\pi/2$ in first quadrant of the plane when the applied  $\mathbf{E}$ is along the tilt axis and it also shows the $\sin \theta$ angular dependence at finite magnetic field for the same configuration. However, the B-dependence of both LMC and PHC is quadratic in B when the electric field is applied perpendicular to the tilt direction. 
The appropriate system for measurement of this anisotropy are recently discovered type II WSMs such as WTe$_{2}$~\cite{Wu_2016} and MoTe$_{2}$~\cite{Deng_2016}.

\textit{Conclusion:} In this work we have presented a quasi-classical theory of chiral anomaly induced planar Hall effect in Weyl semimetals.
We derived an analytical expression for planar Hall conductivity and also elucidated its generic behavior for type-I and type-II WSMs. Unlike anomalous Hall effect \cite{Jung}, to the best of our knowledge PHE has not been described as a topological response function in terms of Berry phases, and our unified treatment of PHE and LMC in terms of chiral anomaly and Berry phase effects, together with experimental predictions in type-I and type-II Weyl semimetals, is an important first step in this direction.

\textit{Acknowledgment:}
The authors (SN and AT) acknowledge the computing facility from DST-Fund for S and T infrastructure (phase-II) Project installed in the Department of Physics, IIT Kharagpur, India.
ST acknowledges support from ARO Grant No: (W911NF-16-1-0182).

\end{document}